\documentclass[conference]{IEEEtran}
\IEEEoverridecommandlockouts
\PassOptionsToPackage{hyphens}{url}\usepackage{hyperref}
\usepackage{url}
\usepackage{cite}
\usepackage{amsmath,amssymb,amsfonts}
\usepackage{algorithmic}
\usepackage{multirow}
\usepackage{graphicx}
\usepackage{textcomp}
\usepackage{xcolor}
\usepackage{subcaption}
\usepackage{tcolorbox}
\usepackage{hyperref}
\def\BibTeX{{\rm B\kern-.05em{\sc i\kern-.025em b}\kern-.08em
    T\kern-.1667em\lower.7ex\hbox{E}\kern-.125emX}}
\begin{document}

\title{Can LLMs Replace Humans During Code Chunking?}

\author{\IEEEauthorblockN{Christopher Glasz, Emily Escamilla, Dr.\ Eric O.\ Scott, Anand Patel, Jacob Zimmer, Colin Diggs, \\Michael Doyle, Dr. Scott Rosen, Dr. Nitin Naik, Dr. Justin F. Brunelle, Dr. Samruddhi Thaker,\\Parthav Poudel, Arun Sridharan, Amit Madan, Doug Wendt, William Macke, and Thomas Schill\textsuperscript{\textsection}}
\IEEEauthorblockA{\textit{The MITRE Corporation} \\
McLean, VA\\
cglasz@mitre.org}
}

\maketitle

\begingroup\renewcommand\thefootnote{\textsection}
\footnotetext{Due to the size of the team, we list everyone here but see the first author's contact information below as the corresponding author.}
\endgroup

\begin{abstract}
Large language models (LLMs) have become essential tools in computer science, especially for tasks involving code understanding and generation. However, existing work does not address many of the unique challenges presented by code written for government applications. In particular, government enterprise software is often written in legacy languages like MUMPS or assembly language code (ALC) and the overall token lengths of these systems exceed the context window size for current commercially available LLMs. Additionally, LLMs are primarily trained on modern software languages and have undergone limited testing with legacy languages, making their ability to understand legacy languages unknown and, hence, an area for empirical study. This paper examines the application of LLMs in the modernization of legacy government code written in ALC and MUMPS, addressing the challenges of input limitations. We investigate various code-chunking methods to optimize the generation of summary module comments for legacy code files, evaluating the impact of code-chunking methods on the quality of documentation produced by different LLMs, including GPT-4o, Claude 3 Sonnet, Mixtral, and Llama 3. Our results indicate that LLMs can select partition points closely aligned with human expert partitioning. We also find that chunking approaches have significant impact on downstream tasks such as documentation generation. LLM-created partitions produce comments that are up to 20\% more factual and up to 10\% more useful than when humans create partitions. Therefore, we conclude that LLMs can be used as suitable replacements for human partitioning of large codebases during LLM-aided modernization.
\end{abstract}


\section{Introduction}
\label{sec:intro}

Large language models (LLMs) have become pivotal in computer science due to their ability to understand and generate code for modern programming languages \cite{chen2021evaluating,vaithilingam2022expectation,hou2024large,jiang2024survey}. Code modernization, the process of converting legacy systems written in legacy languages into modern languages, has emerged as an application for LLMs due the incredible amount of time and expertise required for modernization. Extensive research has been done in the area of optimizing LLM performance for specific tasks through techniques like prompt engineering, which involves manipulating task instructions to enhance output quality \cite{wang2024enhancing,taherkhani2024epic,zhang2024instruct}. However, there remains a significant gap in understanding the impact of input sizes of code partitions on the quality of LLM software documentation outputs, especially when handling large volumes of code such as those in government legacy IT systems.

For tasks involving code, it is common practice to provide an entire file to an LLM for context, operating under the assumption that more context leads to better output quality. However, while the context windows of LLMs continue to grow, legacy language source files often exceed these limits. Government legacy systems are especially large with millions of lines of code. These instances necessitate ``chunking," a process by which users manually or programmatically break up or partition longer documents into smaller pieces to facilitate use in LLM software modernization tasks. 

Unlike natural language inputs, software follows a highly structured format and often contains critical dependencies, which makes chunking a delicate operation. For example, splitting code in the middle of a logic routine such as a function definition or control-flow structure would separate valuable context and may severely hamper an LLM's ability to understand the code. Therefore, chunking a source file strictly by lines of code may divide essential information across multiple chunks. While abstract syntax tree (AST) parsers can aid in chunking code according to its structure, ASTs are not available for all programming languages and creating an AST is a non-trivial and time-consuming undertaking. These limitations make LLM Partitioning an appealing strategy for modernization tasks. Understanding these dynamics is crucial for LLM optimization to maximize output quality, minimize hallucinations, and save computational and human engineering time.

Due to the limited availability of subject-matter experts (SMEs) in legacy programming languages, modernization efforts increasingly rely on LLMs to understand legacy code and aid in modernization. In this context, generating documentation becomes a crucial application of LLMs, as comments play a vital role in ensuring code conversion accuracy \cite{diggs2024leveraging}. In this study, we evaluate LLM performance on IBM mainframe assembly language code (ALC) and MUMPS languages, two legacy programming languages that are relatively obscure in LLM training datasets but remain critical to government systems. To support modernization of these legacy systems, LLM-generated documentation must be both reliable and of high quality. Building on this foundation, this study investigates two primary research questions: 

\begin{itemize}
\item \textbf{RQ1: What automated chunking method most closely resembles human partitioning?}
\item \textbf{RQ2: What chunking method results in the highest quality of generated documentation?}
\end{itemize}

By exploring these questions, we aim to identify ways to improve LLM performance and reliability in tasks that impact the efficiency and accuracy of legacy code modernization, such as software documentation.

\section{Related Work}
\label{sec:related}


Much of the recent innovation and performance improvements offered by LLMs have been centered around increasing context window lengths. Researchers have combined pre-training methods like positional embeddings (especially RoPe \cite{su2024roformer} and its extensions \cite{zhang2024extending}) with long-document training datasets (ex. LongBench \cite{bai2023longbench}) to enable transformer architectures that can process long-term dependencies among documents of hundreds of thousands of tokens in length or more.

In general, LLM evaluation studies show that these models are able to recall and leverage diverse informational content in very long documents \cite{laban2024summary}.  For particular operational tasks, however, it is not clear that simply feeding LLMs as much context information as possible for the task at hand is always beneficial.  In some cases, researchers have demonstrated that LLMs are strongly biased toward contextual information they receive, excluding parametric knowledge that may facilitate generalization \cite{wadhwa2024rags}.  A key design decision for many applications, then, is determining what strategy best processes a corpus of long documents into smaller chunks for processing by an LLM.

Researchers have studied the chunking problem most extensively in the domain of retrieval-augmented generation (RAG). In RAG systems, developers must balance breaking documents into small pieces to retrieve specific information while retaining enough context to identify documents as relevant to a query. The first RAG systems relied on fixed-length chunking of the input corpus \cite{lewis2020retrieval}, but the most common methods today employ semantic similarity as well as recursive splitting with an objective of keeping sentences and paragraphs together \cite{finardi2024chronicles,qu2024semantic}.\footnote{See for example LangChain's influential \href{https://python.langchain.com/v0.1/docs/modules/data_connection/document_transformers/recursive_text_splitter/}{RecursiveCharacterTextSplitter} and \href{https://python.langchain.com/docs/how_to/semantic-chunker/}{SemanticChunker} APIs, and LlamaIndex's \href{https://docs.llamaindex.ai/en/stable/examples/node_parsers/semantic_chunking/}{SemanticSplitterNodeParser}.} Many of these approaches use a ``small-to-big" strategy, beginning with a small portion of text and then iteratively adding adjacent regions that fall within a similarity threshold \cite{gao2023retrieval}. More recent chunking strategies, like meta-chunking \cite{zhao2024meta}, extend semantic approaches using an LLM to determine segmentation points \cite{duarte2024lumberchunker}, or forgoing chunking entirely through chunking-free in-context retrieval \cite{qian2024grounding}.

The effect of chunking strategies on software-related tasks, by contrast, has rarely been studied. Previous publications involving experimentation of code translation or summarization use a fixed-size chunking strategy or adopt a policy of using one chunk for each file in a commit \cite{cordeiro2024empirical}. Zhou et al. \cite{zhou2024llm} use LangChain's built-in \href{https://js.langchain.com/v0.1/docs/modules/chains/document/map_reduce/}{MapReduce-style}\footnote{\url{https://js.langchain.com/v0.1/docs/modules/chains/document/map_reduce/}} methods of splitting and aggregating chunks for large Python documents in code understanding tasks. For tasks that take code as input, syntactic approaches to chunking are a natural choice. For instance, Koziolek et al. \cite{koziolek2024llm} chunk code into function blocks for RAG-based code generation and argue that this meets the intuition that it is a best practice to ``create chunks with semantically related information''. Wang at al. \cite{wang2024python}, meanwhile, contrast chunking by lines with chunking by condition statements when generating Z3 code from Python with the assistance of RAG. 

In particular, optimal code chunking strategies for comment generation remains a research gap. In the context of modernization, ensuring the generation of high quality documentation by LLMs is paramount. While researchers are investigating the use of LLMs for direct code translation, many challenges surrounding LLM direct translation for legacy languages remain unsolved \cite{eniser2024towards,pietrini2024bridging}. Therefore, in lieu of directly translating legacy code with LLMs, LLMs can generate code documentation, like comments, that aid human developers in code translation tasks \cite{diggs2024leveraging}. Comment generation is a potentially context-heavy task, as developers often report that comments that simply recapitulate what a single line of code performs are less useful than comments that describe code in the wider context of the function, module, or application. Therefore, it is beneficial to provide as much source code as possible to an LLM when performing summarization tasks. Where context windows impose limitations and require code chunking, it is important to understand the impact of and identify optimal code chunking strategies. 

Based on the existing research, we hypothesize that \textbf{LLM Partitioning will most closely resemble human partitioning} as opposed to rule or size based partitioning methods. We also hypothesize that \textbf{LLM Partitioning with no context limit will result in the highest quality of generated comments}, following the logic that additional context will provide more useful comments. 

\section{Methods}
\label{sec:methods}


In this section, we outline the methodologies we used to investigate input limitations for LLMs and the impact of partitioning strategies on the quality of code documentation generated by LLMs. We aim to identify optimal partitioning strategies that enhance the generation of quality comments. The following subsections detail the partitioning methods, comment generation process, and evaluation methods used in our study.

\subsection{Partitioning Methods}
\label{sec:partition}
To understand the effect of chunking on generation quality, we experimented with three categories of partitioning strategies with two chunking methods per category: naïve chunking, structure-based chunking, and expert-driven chunking. The categories and associated chunking methods are shown in Table \ref{tab:method_table}.

\begin{table}[]
\caption{Breakdown of the chunking methods by type including the context windows tested in this experiment.}
\resizebox{\columnwidth}{!}{%
\begin{tabular}{|l|l|l|}
\hline
Category                         & Chunking Method     & Context Window         \\ \hline
\multirow{2}{*}{Naive Chunking}  & Full File           & Unlimited              \\ \cline{2-3} 
                                 & Fixed Length Chunks & Variable               \\ \hline
\multirow{2}{*}{Structure-Based} & Single Module       & Unlimited              \\ \cline{2-3} 
                                 & Multiple Modules    & Variable               \\ \hline
\multirow{2}{*}{Expert-Driven}   & Human Partitions    & Unlimited              \\ \cline{2-3} 
                                 & LLM Partitions      & Variable and Unlimited \\ \hline
\end{tabular}%
}
\label{tab:method_table}
\end{table}

For the naïve chunking strategy, we used \textbf{Full File} and \textbf{Fixed Length Chunking} methods. With \textbf{Full File} chunking, the code is provided to the LLM in its entirety. \textbf{Fixed Length Chunking} splits the code at line breaks without exceeding the defined token limit. However, arbitrary line breaks may result in language structures like functions or loops being curtailed, resulting in critical logic being omitted from chunks.

To take the code structure into consideration, we employed structure-based chunking with the \textbf{Single Module} and \textbf{Multiple Module} methods. For both methods, we used a language-specific abstract syntax tree (AST) parser to identify complete ``modules'' in the code (i.e., subroutines in MUMPS, CSECTs or DSECTs in ALC). With \textbf{Single Module} chunking, each chunk contains only one module. With \textbf{Multiple Module} chunking, adjacent modules are iteratively merged until they reach the defined token limit. 

Lastly, we employed two expert-driven chunking methods: \textbf{Human Partitions} and \textbf{LLM Partitions}. The SMEs that we asked to create our \textbf{Human Partitions} were programmers experienced with the languages used in experimentation\footnote{Two MUMPS programmers with more than 5 years of combined MUMPS experience, and one ALC programmer with over 40 years of experience with assembly programming.}. Both the human  SMEs and the LLM were tasked to partition code into ``logical blocks that are relatively self-contained and ideally constitute a complete subroutine''\footnote{Partitioning prompt provided in Appendix \ref{sec:partition_prompt}}. Additionally, we provided the LLM with instructions for formatting output as a JSON object and a sample input-output pair. For the \textbf{LLM Partitions} approach, we allowed for the specification of a token limit -- when the model produced chunks that exceeded this limit, we re-queried the LLM with the original input, its partitioning, the list of chunks that exceeded the limit, a multiplier indicating by how much the limit was exceeded, and instructions to shorten those chunks.

For each of the approaches that support variable token limits (\textbf{Fixed Length Chunks}, \textbf{Multiple Modules}, and \textbf{LLM Partitions}), we experimented with four token limits: 512, 1,024, 2,048, and 4,096. Note that we used the GPT-4o tokenizer to count chunk tokens regardless of the model being queried to avoid differences in inputs between experiments. We selected this range for practical reasons: LLMs struggled to partition code from one of our datasets into chunks any shorter than 512 tokens, and in the other dataset, the longest file contained only 4,785 tokens.

\subsection{Data Preparation}
We selected two codebases--one written in ALC and the other in MUMPS--to test our methods. For the ALC corpus, we collected the code from Walmart Labs' open source software for file access management\footnote{\url{https://github.com/walmartlabs/zFAM}} and unique identifier generation\footnote{\url{https://github.com/walmartlabs/zUID}}. The MUMPS corpus consists of the Incomplete Records Tracking (IRT) module of the Open Source Electronic Health Record Alliance (OSEHRA) VistA open source electronic health record (EHR) software\footnote{\url{https://github.com/WorldVistA/VistA}}, which has been utilized by the U.S. Department of Veterans Affairs. We selected the IRT module as being representative of the overall complexity of the VistA codebase. Table \ref{tab:data} provides some statistics of the ALC and MUMPS corpora. We observed repeatable behavior of our methods, with government systems being similar to the Walmart and IRT code. We evaluate our methods on these open-source corpora because we cannot release the findings from our work on government systems due to sensitivities and releasability restrictions.

\begin{table}[htbp]
\caption{Basic statistics on the two legacy code datasets investigated.}
\begin{center}
\begin{tabular}{|lrrrrr|}
    \hline
    Language  & Files  & Modules  &  Lines  & Tokens    & Characters   \\ \hline
    ALC       & 12     & 54       &  13,344 & 141,526   & 566,104     \\
    MUMPS     & 78     & 1,948    &  5,107  & 139,117   & 556,468     \\ \hline
    \end{tabular}
\label{tab:data}
\end{center}
\end{table}

To prepare the corpora for comment generation (Section \ref{sec:com_gen}), we stripped each source file of any existing comments and remarks\footnote{For the \textbf{Human Partitions} and \textbf{LLM Partitions} chunking methods, we provided the code to the SMEs and LLMs with comments intact. We then removed the comments from the partitioned outputs.}, and then annotated with module markers. We gave each module marker a unique 8-character random identifier, and inserted it on a new line preceding each module -- these modules correspond directly with modules use in the structure-based chunking methods. For each file we processed, we partitioned it according to our 16 chunking methods: 3 fixed-length approaches (\textbf{Full File}, \textbf{Single Module}, and \textbf{Human Partitions}), 3 variable-limit methods with 4 limits each (\textbf{Fixed-Length Chunks}, \textbf{Multiple Modules}, \textbf{LLM Partitions}), in addition to the \textbf{LLM Partitions} approach with no limit imposed. In total, our corpus contained 16 partitioned files for each original file.

\subsection{Comment Generation}
\label{sec:com_gen}

To measure the impact of partitioning, we assessed the quality of the LLM output for a given task across multiple LLMs. Specifically, we are interested in the impact of code partitioning on the quality of documentation generation. For this portion of the experiment, we prompted the four LLMs (GPT-4o, Claude 3 Sonnet, Mixtral, and Llama 3) to generate module-level inline comments. To do this, we prompted the LLMs to provide a short documentation string corresponding to each of the uniquely identified module markers that we inserted into the source code\footnote{Comment generation prompt provided in Appendix \ref{sec:doc_prompt}}. 

\subsection{Evaluation Methods}

We evaluated the partitioning methods across two categories of measures: \textbf{quality of the partitioning method} and \textbf{quality of LLM-generated documentation} using the partitions. These measures allow us to assess the validity of utilizing automated partitioning to generate trustworthy and high quality documentation for modernization.

\subsubsection{Measuring the Quality of the Partitioning Method}

For this evaluation, we used human partitioning as our ground truth for the logical partitioning of code files. We used the metrics recall and precision to quantify the similarity between human partitioning and other chunking methods. For this study, recall is computed as the fraction of ground truth split points that were identified by the chunking method and precision is the fraction of split points identified by the chunking method that were ground truth split points. For each chunking method and chunking size, we calculated the recall and precision of the partition in relation to the SME partition. We used GPT-4o partitions as a representative sample of the LLM Partitions chunking methods for this aspect of the experiment. 

\subsubsection{Measuring the Quality of LLM-Generated Documentation}
\label{sec:self_eval}

While similarity to human partitioning provides insight for automated implementations, assessing the quality of LLM-generated documentation is the final judge of partitioning method quality. To measure the quality of generated documentation, we prompted GPT-4o and Claude 3 Sonnet to evaluate the comments generated with the methodology provided in Section \ref{sec:com_gen}. We quantified quality through evaluation metrics characterizing \textit{completeness}, \textit{hallucination}, \textit{readability}, and \textit{usefulness} and provided the LLMs with guidelines for each characteristic \footnote{Evaluation prompt and metric definitions provided in Appendix \ref{sec:eval_prompt}}. 

The approach we employed for generating LLM evaluations of module summaries was similar to the one used for generating those summaries. From the prepared code with original comments removed and module markers added, we inserted the generated comments directly after their corresponding marker in the code. We then provided the LLM with the entire file and prompted it to generate a JSON-formatted output mapping the unique module identifiers to evaluation objects containing scores for the four characteristics described above\footnote{Evaluation prompt provided in Appendix \ref{sec:eval_prompt}}. 

Previous work studying the effects of batch prompting\cite{cheng2023batchprompting} suggest that performance may degrade when the input prompt length is not significantly greater than the output length. We therefore limit the number of comments graded simultaneously per request by including only a connected subset of five comments in each prompt, and repeating the task with different ``windows'' of comments until all generations have been graded.

We initially used a four-point grading scale for each of these characteristics, to mirror the rubric we offered human SMEs in other internal experiments. However, in initial experimentation, we found that LLMs provided more accurate scoring when not restricted to a 4-point scale and that LLMs tend to score favorably if not instructed to grade more critically. To bias the LLMs toward lower scores, we modified our prompting strategy to request a score for each characteristic on a 100-point scale with the inclusion of a reasoning for the score and instructed the LLM to be a ``hard grader''. 

To obtain robust and easily-visualized results, we generated 10 independent evaluations of each comment and took the mean of the scores. We observed intra-class correlation coefficients for average random raters (ICC2k) of 0.960 on the MUMPS dataset and 0.908 on the ALC dataset, suggesting that there's a high level of agreement between iterations of GPT-4o evaluations.

In total, this evaluation required 1.28 million requests to GPT-4o (2,002 module comments generated by 4 models using 16 chunking methods and 10x coverage), taking 37 days and costing \$8,547.

\section{Results}
\label{sec:results}

This section presents the findings of our study, organized to address the two research questions (\textbf{RQ1} and \textbf{RQ2}) outlined in the \hyperref[sec:intro]{Introduction}. First, we evaluate the quality of the partitioning methods in aligning with SME-defined partitions across two code corpora: MUMPS and ALC. Next, we assess the quality of LLM-generated documentation based on the partitioning methods and models. Finally, we analyze the impact of context limits on partitioning method performance, focusing on the metrics of hallucination and usefulness. The results highlight the effectiveness of the \textbf{LLM Partitioning} method in aligning with \textbf{Human Partitioning} and generating high-quality documentation, while also exploring the variability in performance across models and corpora.

\subsection{Quality of the Partitioning Methods}
\begin{figure*}[t]
\centerline{
    \begin{subfigure}{.5\textwidth}
      \centering
      \includegraphics[width=\linewidth]{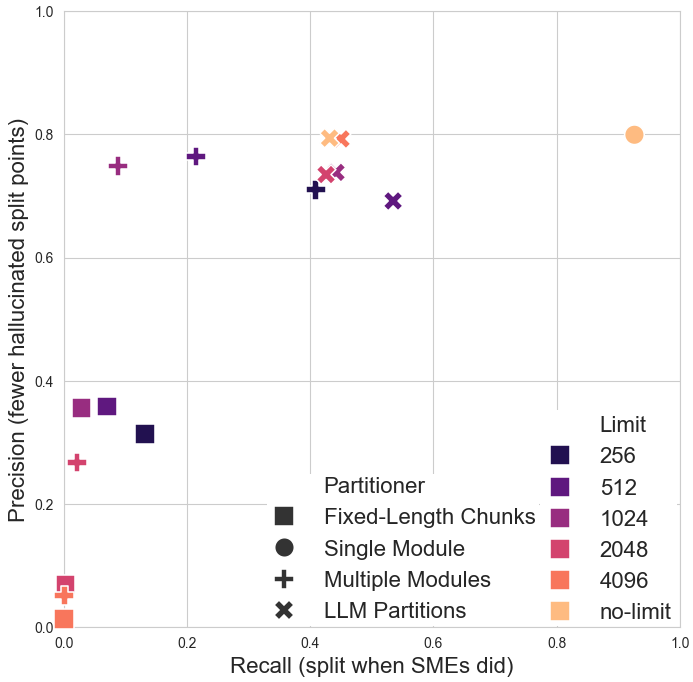}
      \caption{Partition performance on MUMPS corpus}
      \label{fig:mumps_partitions}
    \end{subfigure}%
    \begin{subfigure}{.5\textwidth}
      \centering
      \includegraphics[width=\linewidth]{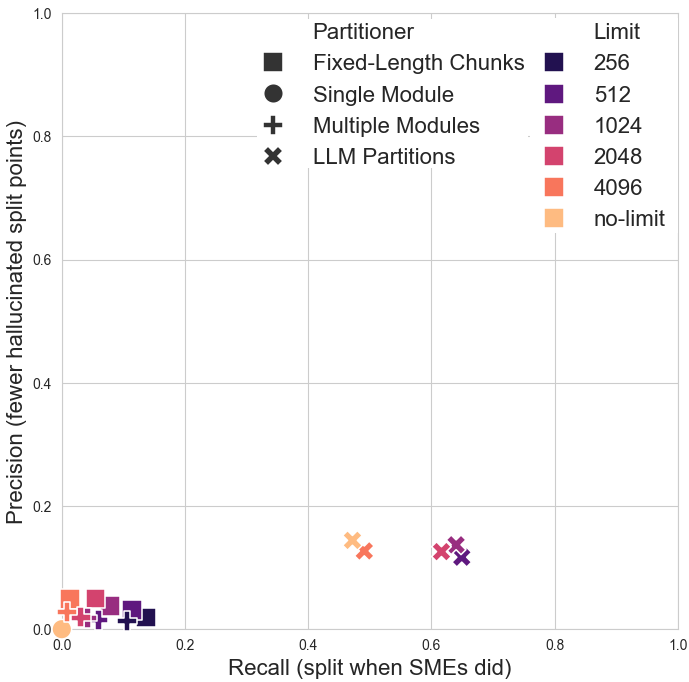}
      \caption{Partition performance on ALC corpus}
      \label{fig:alc_partitions}
    \end{subfigure}
}
\caption{Comparing precision and recall of partitioning methods in relation to SME partitioning on each code corpus. Higher precision and recall values reflect closer alignment with SME partitions. Values close to 0 reflect minimal alignment with SME partitions. For each corpus, LLM Partitions most closely resemble human partitioning.}
\label{fig:partition_performance}
\end{figure*}

To address \textbf{RQ1}, we calculated the precision and recall for every partitioned file in relation to SME partitioning for the MUMPS corpus (Figure \ref{fig:mumps_partitions}) and ALC corpus (Figure \ref{fig:alc_partitions}). As shown by the higher precision and recall in Figure \ref{fig:mumps_partitions}, the partitions for the MUMPS corpus show overall closer alignment with SME partitions than the ALC corpus which demonstrated precision values close to zero and most recall values below 0.2 as shown in Figure \ref{fig:alc_partitions}. With the exception of \textbf{Single Module} chunking, \textbf{LLM Partitions} outperformed all other partitioning methods regardless of partition limit and programming languages. Because MUMPS labelled blocks are very similar to the idea of subroutines, the \textbf{Single Module} method (indicated by the circle icon) partitioned the file at almost all the same points that SMEs did, resulting in near-perfect recall. \textbf{Multiple Modules} chunking has comparable precision with \textbf{Single Module} chunking, but far lower recall, meaning that Single Module chunking correctly identifies more of the human partition points. As expected due to their arbitrary nature, \textbf{Fixed-Length Chunks} have minimal overlap with SME partitions. Both the \textbf{Fixed-Length Chunks} and \textbf{Multiple Modules} approaches have a similar curve in precision-recall space when varying the context limit -- a peak in precision at the 512-token limit, abysmal precision at 4096 tokens, and steadily improving recall as the limit gets smaller. With MUMPS, this pattern makes sense, as most of the source files fit within a 4096-token window, resulting in extremely few partitions with these approaches, while our SMEs inserted partitions between nearly every labeled block.

As shown in Figure \ref{fig:alc_partitions}, matching SME partitions in ALC is generally a much more challenging task than in MUMPS, primarily due to the relative sparsity of SME partition points. With these ``smaller targets'', only \textbf{LLM Partitions} show any resemblance to the SME partitions, while all other methods scored close to zero on both precision and recall. For the ALC corpus, the LLM Partitions chunking method demonstrated recall comparable to that of MUMPS source code, correctly identifying SME partition points at a rate consistent with its performance on the MUMPS corpus. However, it had low precision, similar to other partitioning methods on the ALC corpus, and identified additional partition points not recognized by human partitioning. Overall, in response to \textbf{RQ1}, \textbf{LLM Partitions} exhibit the highest precision and recall in most closely resembling human partitioning for both ALC and MUMPS.

\subsection{Quality of LLM-Generated Documentation}
\begin{figure*}[htbp]
\centerline{\includegraphics[width=\textwidth]{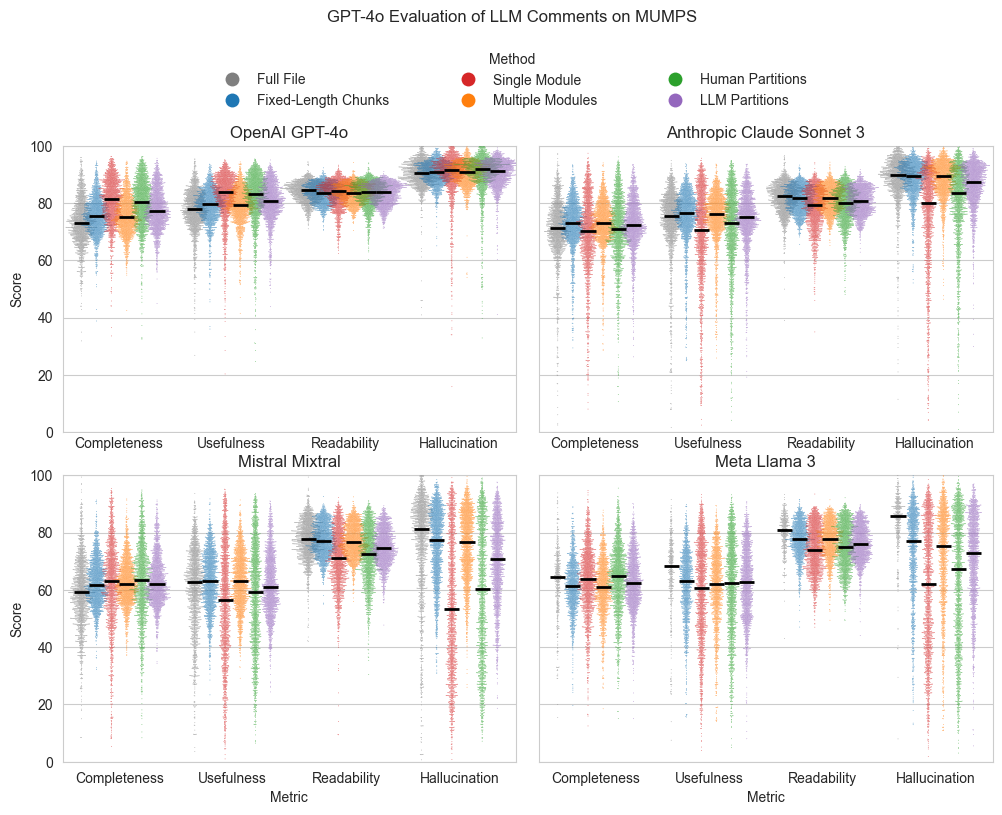}}
    \caption{Completeness, usefulness, readability, and hallucination scores for GPT-4o, Claude Sonnet 3, Mixtral, and Llama 3 for each chunking method on the MUMPS corpus. The evaluated hallucination score is inversely related to the prevalence of hallucination meaning a higher score means fewer hallucinations. Black horizontal lines indicate the median values.}
    \label{fig:no_limits_mumps}
\end{figure*}

\begin{figure*}[htbp]
\centerline{\includegraphics[width=\textwidth]{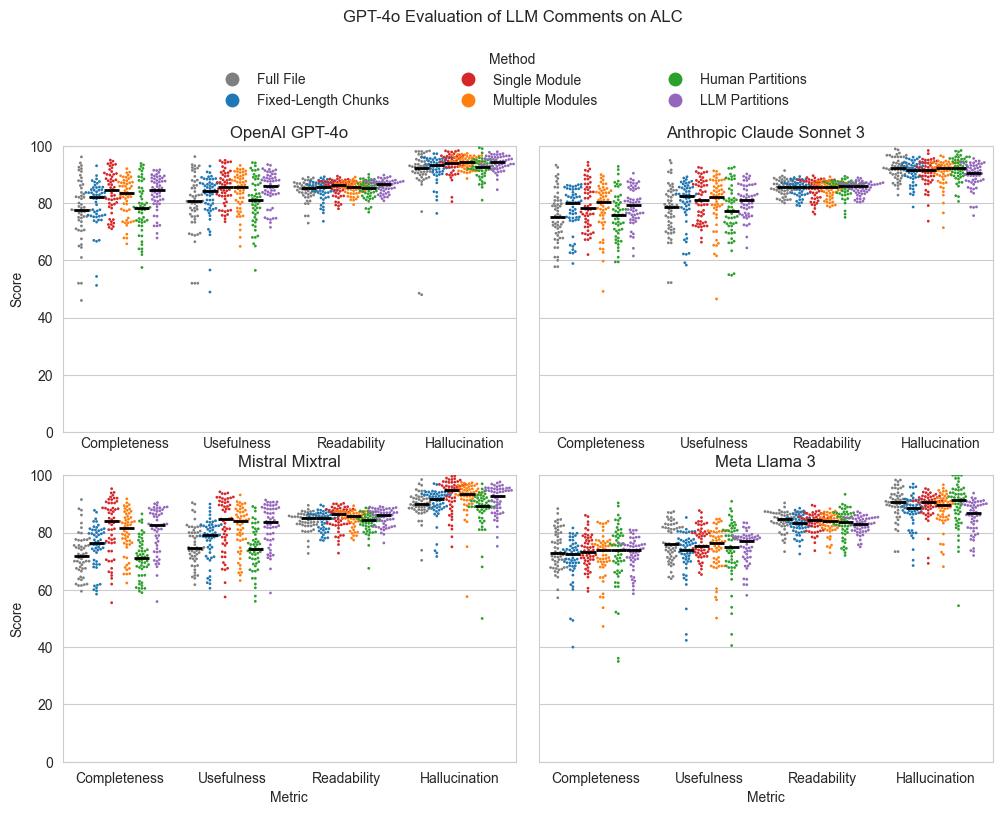}}
    \caption{Completeness, usefulness, readability, and hallucination scores for GPT-4o, Claude Sonnet 3, Mixtral, and Llama 3 for each chunking method on the ALC corpus. The evaluated hallucination score is inversely related to the prevalence of hallucination meaning a higher score means fewer hallucinations. Black horizontal lines indicate the median values.}
    \label{fig:no_limits_alc}
\end{figure*}

To address \textbf{RQ2}, we plotted the LLM self-evaluation scores for each partitioning method and LLM with the results for the MUMPS codebase shown in Figure \ref{fig:no_limits_mumps} and the ALC codebase shown in Figure \ref{fig:no_limits_alc}. As shown in Figure \ref{fig:no_limits_mumps}, the performance of partitioning methods on the MUMPS codebase varies across models and metrics. With GPT-4o, for example, \textbf{Single Module} chunking scores higher than the other chunking methods for completeness and usefulness. However, \textbf{Single Module} chunking is one of the lowest scoring chunking methods across all metrics for Claude 3 Sonnet, Mixtral, and Llama 3. \textbf{LLM Partitioning} performs similarly to \textbf{Human Partitioning} across all four models for the metrics of completeness, usefulness, and readability. However, comments generated from \textbf{LLM Partitions} are notably scored as more factual than comments generated from \textbf{Human Partitions} for Claude 3 Sonnet, Mixtral, and Llama 3. For the MUMPS codebase, \textbf{LLM Partitions} produced comments that were up to 10\% more factual than \textbf{Human Partitions}. The improved performance of \textbf{LLM Partitioning} over \textbf{Human Partitioning} is especially pronounced with the ALC codebase. Across all four LLMs, \textbf{LLM Partitions} are among the highest scores for the metrics of usefulness and completeness. Specifically, \textbf{LLM Partitions} out-perform \textbf{Human Partitions} by up to 10\% in usefulness and completeness for the ALC codebase.

\subsection{Chunking Method Performance Across Context Limits}
\begin{figure*}[htbp]
\centerline{\includegraphics[width=\textwidth]{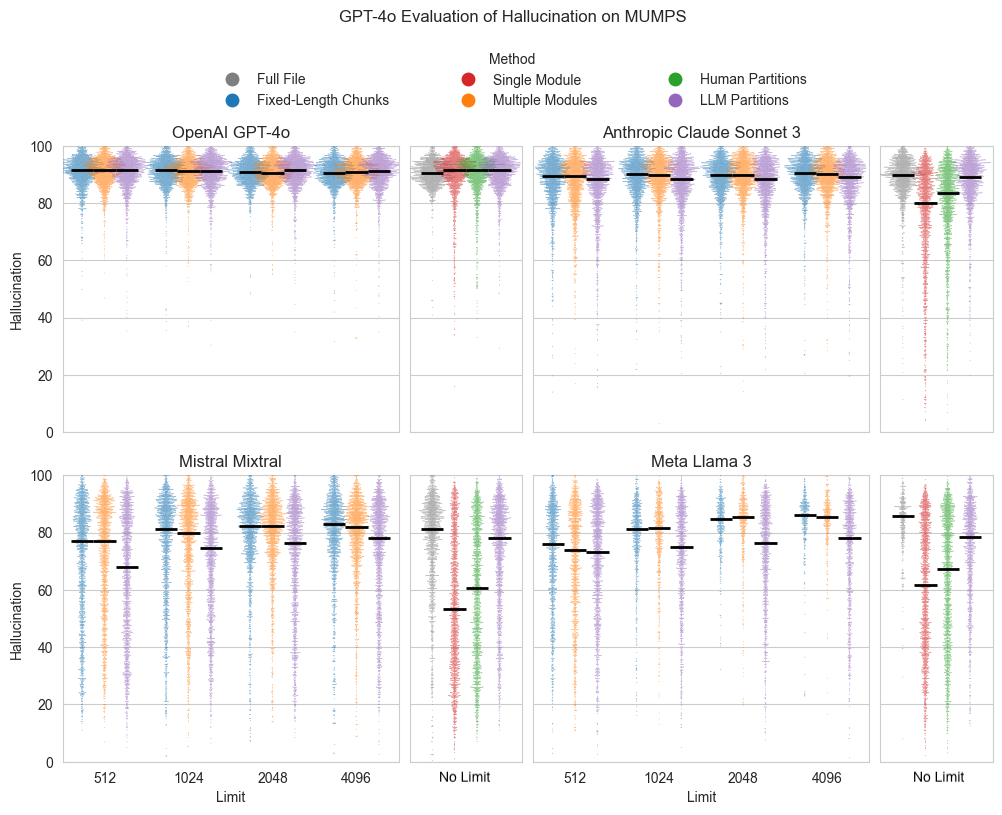}}
    \caption{Hallucination scores for all generated comments for the MUMPS corpus based on partitioning method, partitioning limit, and the LLM that generated the comment. The evaluated hallucination score is inversely related to the prevalence of hallucination meaning a higher score means fewer hallucinations. Black horizontal lines indicate the median values. For chunking methods with variable context lengths, the token limit is on the x-axis. Methods without variable context lengths are presented in a separate box to the right for each model.}
    \label{fig:hallucination_swarm_mumps}
\end{figure*}

\begin{figure*}[htbp]
\centerline{\includegraphics[width=\textwidth]{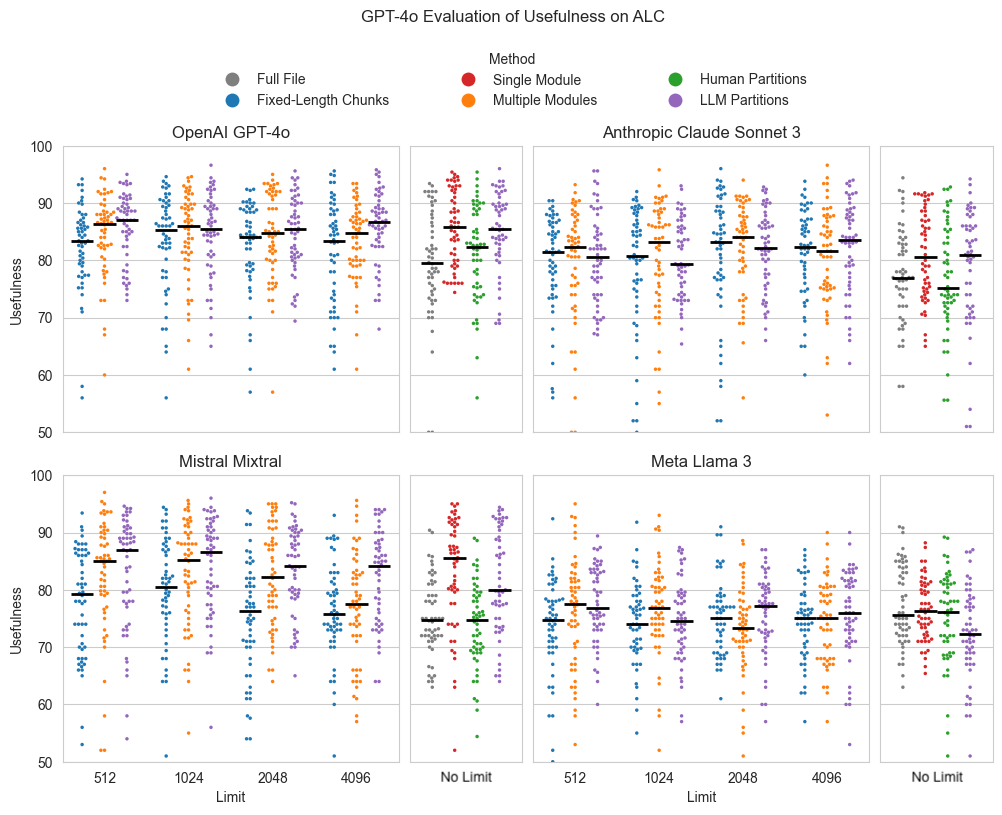}}
    \caption{Usefulness scores for all generated comments for the ALC corpus based on partitioning method, partitioning limit, and the LLM that generated the comment. Black horizontal lines indicate the median values. For chunking methods with variable context lengths, the token limit is on the x-axis. Methods without variable context lengths are presented in a separate box to the right for each model.}
    \label{fig:usefulness_swarm_alc}
\end{figure*}

Next, we assessed the impact of context limits on the performance of partitioning methods with respect to comment factualness (Figure \ref{fig:hallucination_swarm_mumps}) and usefulness (Figure \ref{fig:usefulness_swarm_alc}). Figure \ref{fig:hallucination_swarm_mumps} shows hallucination scores for each model (quadrant) and chunking method (color) on the MUMPS corpus. For chunking methods with variable context lengths (i.e., fixed-length chunks, multiple models), the token limit is on the x-axis and methods where no such limit is imposed (i.e., Full File, Single Module) are presented in a separate box to the right for each model.

As shown by Figure \ref{fig:hallucination_swarm_mumps}, chunking methods have a greater impact on LLM performance for Mixtral and Llama 3 than for GPT-4o and Claude Sonnet 3. GPT-4o and Claude Sonnet 3 achieved consistently high factualness with GPT-4o demonstrating comparable results across chunking methods. Claude Sonnet 3 comments had slightly more variance between chunking methods with \textbf{Single Module} chunking producing the worst hallucination scores followed by \textbf{Human Partitions}. Mixtral and Llama 3 achieved median hallucination scores 20-40 points lower than GPT-4o and Claude Sonnet 3 and had significantly higher variance. The effect of chunking methods is stark in these poorer-performing LLMs. Notably, \textbf{LLM Partitions}, regardless of context limit, produce comments that are up to 20\% more factual due to reduced hallucination than \textbf{Human Partitions}.

For methods with variable context length, context length is positively correlated with comment accuracy. This suggests that increasing the amount of context available to the model can lead to more accurate comments. Supporting this, the \textbf{Single Module} chunking method, which provides the most limited context, produced the least factual comments by far. As context increased, the comments produced from the \textbf{Fixed-Length Chunks} and \textbf{Multiple Modules} methods increased in factualness with similar performance to each other across all four token limits and all four LLMs. This trend is likely driven by the fact that MUMPS is a highly terse language with small subroutines (around 72 tokens); therefore, with even a 256-token limit, most windows will include multiple adjacent subroutines. 

The positive correlation between context length and comment accuracy seen with the MUMPS corpus ($\rho$=0.199, p=8.80e-10) does not hold in the ALC corpus. For the ALC corpus, all LLMs and chunking methods produced high scores (means of 90 or higher) on hallucination, and the little variation that occurred indicated that larger contexts were \textit{detrimental} to accuracy, rather than beneficial as previously assumed. This was consistent across metrics, as presented in Figure \ref{fig:usefulness_swarm_alc}, which shows findings on the usefulness of comments generated for our ALC corpus. Here, the \textbf{Full File} approach consistently produced the lowest scores, while the \textbf{Single Module} and \textbf{LLM Partitions} methods performed best.

As shown in Figure \ref{fig:usefulness_swarm_alc}, for the ALC corpus, chunking method and context limit impacted model performance differently across each of the four models. Both Mixtral and Llama 3 produce more useful comments when provided chunks produced with smaller context limits. Mixtral produced the most useful comments with \textbf{LLM Partitions} and the lowest context limit of 512 tokens followed by the \textbf{Single Module} method. The inverse is true for Claude Sonnet 3 which produces more useful comments with chunking methods with higher context limits, and \textbf{LLM Partitions} with the highest (but not unlimited) context limit of 4096 tokens resulted in the highest median usefulness. However, \textbf{Human Partitions} and \textbf{Full File} methods with no context limit produce the lowest usefulness scores for Claude Sonnet 3. For GPT-4o, the \textbf{LLM Partitions} and \textbf{Single Module} methods achieve the highest median usefulness scores and perform similarly across context limits. The variability of model performance across chunking method and context limit indicate that there may not be a once-size-fits-all chunking method appropriate for all models. Despite this variability, \textbf{LLM Partitions} emerge as a front-runner in method performance across most models and metrics. In the ALC corpus, \textbf{LLM Partitions} produce comments that are up to 10\% more useful to human developers than \textbf{Human Partitions}.

For both corpora and nearly all metrics and all four LLMs, the \textbf{LLM Partitions} chunking method performed at least on par, but usually \textit{better than} the \textbf{Human Partitions} approach. Further, \textbf{LLM Partitions} had the highest median score for all LLMs and all metrics other than hallucination on the ALC corpus. In response to \textbf{RQ2}, generally, across all 4 metrics and all 4 LLMs, \textbf{LLM Partitioning} results in the most consistently high quality documentation. While LLM Partitioning did not always produce the highest score, it results in consistently high scores without the large swings in performance found in other chunking methods depending on language and model. Therefore, LLM Partitioning is a viable strategy for partitioning large code bases, like those found in government legacy IT systems. LLM Partitioning is a way to reduce human cost and effort during LLM-aided modernization as well as increase LLM performance in downstream tasks such as comment generation.

\section{Discussion}
\label{sec:discussion}


When assessing \textbf{RQ1}, chunking methods as a proxy for SME partitioning, we identified that the \textbf{LLM Partitions} approach aligns most closely with \textbf{Human Partitions} for both ALC and MUMPS files. Additionally, while the \textbf{Single Module} chunking method performs well for MUMPS files, AST parsers are often not available for legacy languages, and developing an AST or rule-based parser may be prohibitively costly. Therefore, LLM Partitioning is the most viable method.

Regarding \textbf{RQ2}, documentation quality varied significantly across models and partitioning methods, with GPT-4o performing the best on average and Mixtral and Llama 3 scoring the poorest. Chunking method had significant impacts on downstream documentation quality, but that impact varied in strength between corpora. We observed a positive correlation between chunk size and quality for metrics on MUMPS inputs, while we observed no statistically significant correlation between context limit and quality metrics for ALC. Models also responded to chunking methods differently. GPT-4o produced its most useful and complete comments when provided single MUMPS subroutines, while this was the least effective chunking approach for Claude Sonnet 3, Mixtral, and Llama 3. 


\section{Conclusion}
\label{sec:conclusion}


This study explored the impact of code chunking methods on LLMs in the context of legacy code modernization, focusing on downstream documentation generation quality. We presented an approach to partitioning input code using LLMs, and our findings indicate that the resulting chunks can approximate human intuition on code partitioning, making it a viable method for chunking legacy code when ASTs are unavailable or costly to develop. Among the LLMs tested, GPT-4o demonstrated superior performance, particularly excelling in scenarios with constrained information, which may indicate that it is relying on semantic information to ``hallucinate'' accurate documentation.

The results also reveal that partitioning methods do not universally impact all LLMs in the same way. Therefore, the choice of partitioning strategy should be tailored to the specific model being used. This insight is crucial for optimizing LLM performance in code modernization tasks, ensuring that the generated documentation is both reliable and of high quality. Overall, across all 4 metrics and all 4 LLMs, LLM Partitioning results in the most consistently high quality documentation. LLM-created partitions produce comments that are up to 20\% more factual and up to 10\% more useful to human developers than when humans create partitions. Therefore, we conclude that legacy modernization would achieve greater downstream output quality with the utilization of high-performing LLMs to partition code. We evaluated our methods on the publicly available ALC and MUMPS corpora since we are unable to publish findings from our work related to government systems. However, our researchers have observed consistent behavior in our methods pertaining to RQ2, noting that government systems exhibit similarities to the corpora used in this experiment. This finding helps automate one component of LLM-aided modernization and reduce overall human cost of code preparation. Overall, this research contributes to a deeper understanding of how LLMs can be effectively utilized in the modernization of legacy systems, paving the way for more efficient and accurate code transformation processes. 

\section*{Acknowledgment}

This work was supported by the MITRE Independent Research \& Development program. \copyright2025 The MITRE Corporation. All Rights Reserved. Approved for Public Release; Distribution Unlimited. Public Release Case Number 25-1251.

\bibliographystyle{IEEEtran}
\bibliography{main}

\begin{thebibliography}{10}
\providecommand{\url}[1]{#1}
\csname url@samestyle\endcsname
\providecommand{\newblock}{\relax}
\providecommand{\bibinfo}[2]{#2}
\providecommand{\BIBentrySTDinterwordspacing}{\spaceskip=0pt\relax}
\providecommand{\BIBentryALTinterwordstretchfactor}{4}
\providecommand{\BIBentryALTinterwordspacing}{\spaceskip=\fontdimen2\font plus
\BIBentryALTinterwordstretchfactor\fontdimen3\font minus \fontdimen4\font\relax}
\providecommand{\BIBforeignlanguage}[2]{{%
\expandafter\ifx\csname l@#1\endcsname\relax
\typeout{** WARNING: IEEEtran.bst: No hyphenation pattern has been}%
\typeout{** loaded for the language `#1'. Using the pattern for}%
\typeout{** the default language instead.}%
\else
\language=\csname l@#1\endcsname
\fi
#2}}
\providecommand{\BIBdecl}{\relax}
\BIBdecl

\bibitem{chen2021evaluating}
M.~Chen, J.~Tworek, H.~Jun, Q.~Yuan, H.~P. D.~O. Pinto, J.~Kaplan, H.~Edwards, Y.~Burda, N.~Joseph, G.~Brockman \emph{et~al.}, ``Evaluating large language models trained on code,'' \emph{arXiv preprint arXiv:2107.03374}, 2021.

\bibitem{vaithilingam2022expectation}
P.~Vaithilingam, T.~Zhang, and E.~L. Glassman, ``Expectation vs. experience: Evaluating the usability of code generation tools powered by large language models,'' in \emph{Chi conference on human factors in computing systems extended abstracts}, 2022, pp. 1--7.

\bibitem{hou2024large}
X.~Hou, Y.~Zhao, Y.~Liu, Z.~Yang, K.~Wang, L.~Li, X.~Luo, D.~Lo, J.~Grundy, and H.~Wang, ``Large language models for software engineering: A systematic literature review,'' \emph{ACM Transactions on Software Engineering and Methodology}, vol.~33, no.~8, pp. 1--79, 2024.

\bibitem{jiang2024survey}
J.~Jiang, F.~Wang, J.~Shen, S.~Kim, and S.~Kim, ``A survey on large language models for code generation,'' \emph{arXiv preprint arXiv:2406.00515}, 2024.

\bibitem{wang2024enhancing}
T.~Wang, N.~Zhou, and Z.~Chen, ``Enhancing computer programming education with {LLMs}: {A} study on effective {Prompt} engineering for {Python} code generation,'' \emph{arXiv preprint arXiv:2407.05437}, Jul. 2024.

\bibitem{taherkhani2024epic}
H.~Taherkhani, M.~Sepindband, H.~V. Pham, S.~Wang, and H.~Hemmati, ``{EPiC}: {Cost}-effective search-based prompt engineering of {LLMs} for code generation,'' \emph{arXiv preprint arXiv:2408.11198}, Aug. 2024.

\bibitem{zhang2024instruct}
T.~Zhang, Y.~Yu, X.~Mao, S.~Wang, K.~Yang, Y.~Lu, Z.~Zhang, and Y.~Zhao, ``Instruct or interact? {Exploring} and eliciting {LLMs}' capability in code snippet adaptation through prompt engineering,'' \emph{arXiv preprint arXiv:2411.15501}, Nov. 2024.

\bibitem{diggs2024leveraging}
C.~Diggs, M.~Doyle, A.~Madan, S.~Scott, E.~Escamilla \emph{et~al.}, ``Leveraging {LLMs} for legacy code modernization: {Challenges} and opportunities for {LLM}-generated documentation,'' \emph{arXiv preprint arXiv:2411.14971}, Nov. 2024.

\bibitem{su2024roformer}
J.~Su, M.~Ahmed, Y.~Lu, S.~Pan, W.~Bo, and Y.~Liu, ``Roformer: Enhanced transformer with rotary position embedding,'' \emph{Neurocomputing}, vol. 568, p. 127063, 2024.

\bibitem{zhang2024extending}
Y.~Zhang, J.~Li, and P.~Liu, ``Extending llms' context window with 100 samples,'' \emph{arXiv preprint arXiv:2401.07004}, 2024.

\bibitem{bai2023longbench}
Y.~Bai, X.~Lv, J.~Zhang, H.~Lyu, J.~Tang, Z.~Huang, Z.~Du, X.~Liu, A.~Zeng, L.~Hou \emph{et~al.}, ``Longbench: A bilingual, multitask benchmark for long context understanding,'' \emph{arXiv preprint arXiv:2308.14508}, 2023.

\bibitem{laban2024summary}
P.~Laban, A.~R. Fabbri, C.~Xiong, and C.-S. Wu, ``Summary of a haystack: A challenge to long-context llms and rag systems,'' \emph{arXiv preprint arXiv:2407.01370}, 2024.

\bibitem{wadhwa2024rags}
H.~Wadhwa, R.~Seetharaman, S.~Aggarwal, R.~Ghosh, S.~Basu, S.~Srinivasan, W.~Zhao, S.~Chaudhari, and E.~Aghazadeh, ``From rags to rich parameters: Probing how language models utilize external knowledge over parametric information for factual queries,'' \emph{arXiv preprint arXiv:2406.12824}, 2024.

\bibitem{lewis2020retrieval}
P.~Lewis, E.~Perez, A.~Piktus, F.~Petroni, V.~Karpukhin, N.~Goyal, H.~K{\"u}ttler, M.~Lewis, W.-t. Yih, T.~Rockt{\"a}schel \emph{et~al.}, ``Retrieval-augmented generation for knowledge-intensive nlp tasks,'' \emph{Advances in Neural Information Processing Systems}, vol.~33, pp. 9459--9474, 2020.

\bibitem{finardi2024chronicles}
P.~Finardi, L.~Avila, R.~Castaldoni, P.~Gengo, C.~Larcher, M.~Piau, P.~Costa, and V.~Carid{\'a}, ``The chronicles of rag: The retriever, the chunk and the generator,'' \emph{arXiv preprint arXiv:2401.07883}, 2024.

\bibitem{qu2024semantic}
R.~Qu, R.~Tu, and F.~Bao, ``Is semantic chunking worth the computational cost?'' \emph{arXiv preprint arXiv:2410.13070}, 2024.

\bibitem{gao2023retrieval}
Y.~Gao, Y.~Xiong, X.~Gao, K.~Jia, J.~Pan, Y.~Bi, Y.~Dai, J.~Sun, and H.~Wang, ``Retrieval-augmented generation for large language models: A survey,'' \emph{arXiv preprint arXiv:2312.10997}, 2023.

\bibitem{zhao2024meta}
J.~Zhao, Z.~Ji, P.~Qi, S.~Niu, B.~Tang, F.~Xiong, and Z.~Li, ``Meta-chunking: Learning efficient text segmentation via logical perception,'' \emph{arXiv preprint arXiv:2410.12788}, 2024.

\bibitem{duarte2024lumberchunker}
A.~V. Duarte, J.~Marques, M.~Gra{\c{c}}a, M.~Freire, L.~Li, and A.~L. Oliveira, ``Lumberchunker: Long-form narrative document segmentation,'' \emph{arXiv preprint arXiv:2406.17526}, 2024.

\bibitem{qian2024grounding}
H.~Qian, Z.~Liu, K.~Mao, Y.~Zhou, and Z.~Dou, ``Grounding language model with chunking-free in-context retrieval,'' \emph{arXiv preprint arXiv:2402.09760}, 2024.

\bibitem{cordeiro2024empirical}
J.~Cordeiro, S.~Noei, and Y.~Zou, ``An empirical study on the code refactoring capability of large language models,'' \emph{arXiv preprint arXiv:2411.02320}, 2024.

\bibitem{zhou2024llm}
Z.~Zhou, C.~Li, X.~Chen, S.~Wang, Y.~Chao, Z.~Li, H.~Wang, R.~An, Q.~Shi, Z.~Tan, X.~Han, X.~Shi, Z.~Liu, and M.~Sun, ``{LLM}\${\textbackslash}times\${MapReduce}: {Simplified} long-sequence processing using large language models,'' \emph{arXiv preprint arXiv:2410.09342}, Oct. 2024.

\bibitem{koziolek2024llm}
H.~Koziolek, S.~Gr{\"u}ner, R.~Hark, V.~Ashiwal, S.~Linsbauer, and N.~Eskandani, ``Llm-based and retrieval-augmented control code generation,'' in \emph{Proceedings of the 1st International Workshop on Large Language Models for Code}, 2024, pp. 22--29.

\bibitem{wang2024python}
W.~Wang, K.~Liu, A.~R. Chen, G.~Li, Z.~Jin, G.~Huang, and L.~Ma, ``Python symbolic execution with llm-powered code generation,'' \emph{arXiv preprint arXiv:2409.09271}, 2024.

\bibitem{eniser2024towards}
H.~F. Eniser, H.~Zhang, C.~David, M.~Wang, M.~Christakis, B.~Paulsen, J.~Dodds, and D.~Kroening, ``Towards translating real-world code with {LLMs}: {A} study of translating to {Rust},'' \emph{arXiv preprint arXiv:2405.11514}, May 2024.

\bibitem{pietrini2024bridging}
R.~Pietrini, M.~Paolanti, and E.~Frontoni, ``Bridging {Eras}: {Transforming} {Fortran} legacies into {Python} with the power of large language models,'' in \emph{2024 {IEEE} 3rd {International} {Conference} on {Computing} and {Machine} {Intelligence} ({ICMI})}.\hskip 1em plus 0.5em minus 0.4em\relax IEEE, Apr. 2024.

\bibitem{cheng2023batchprompting}
\BIBentryALTinterwordspacing
Z.~Cheng, J.~Kasai, and T.~Yu, ``Batch prompting: Efficient inference with large language model apis,'' 2023. [Online]. Available: \url{https://arxiv.org/abs/2301.08721}
\BIBentrySTDinterwordspacing

\end{thebibliography}

\onecolumn
\begin{appendix}
\section{Appendix}

\subsection{Partition Prompt}
\label{sec:partition_prompt}

\begin{tcolorbox}[colframe=black, colback=white, width=\linewidth, boxrule=0.5mm, arc=0mm, sharp corners]
\ttfamily
    Your purpose is to partition \{SOURCE\_LANGUAGE\} code into self-contained logical blocks.

    Partition the \{SOURCE\_LANGUAGE\} code into logical blocks. Each block should be relatively self-contained and ideally constitute a complete "subroutine", including any associated comments. These breakpoints should usually be inserted between labeled blocks, but perhaps not between *every* labeled block (depending on things like fallthrough).

    INPUT FORMAT:\\
    Each line of code has been prepended with an 8-character unique ID. A Python example would look like this:
    \{EXAMPLE\_INPUT\}

    And your output might look like this:
    \{EXAMPLE\_OUTPUT\}

    You are to output a JSON object containing a subset of these IDs, corresponding to the lines that should start a new block. Each partition should be paired with an explanation (please output the explanation first, before giving the line ID).
    
    Input:\\
    \{SOURCE\_CODE\}
\end{tcolorbox}

\subsection{Documentation Prompt}
\label{sec:doc_prompt}
\begin{tcolorbox}[colframe=black, colback=white, width=\linewidth, boxrule=0.5mm, arc=0mm, sharp corners]
\ttfamily
    
        You are a senior software engineer tasked with documenting \{SOURCE\_LANGUAGE\} code.\\

        The \{SOURCE\_LANGUAGE\} code provided below has commented tags delineating modules. These tags take the form `<MODULE \#>`, where `\#` takes the place of an 8-character alphanumeric ID, and are associated with the code immediately below it.\\

        You are to write explanatory documentation for each of these labeled modules. This documentation should summarize the intended functionality of the code, define any parameters or outputs, and describe any side-effects or exceptions that may arise in its execution. This documentation will be utilized by engineers to maintain and modernize the code, so it is vital that all the code's behavior is captured.\\

        Your response should be formatted as a JSON object, where the keys are the alphanumeric IDs and the values are the documentation strings. Be sure to include entries for ALL placeholders present in the input. Do not provide any other commentary, do not write any code.\\

        Please provide comments for the following code:\\
        \{SOURCE\_CODE\}\\
\end{tcolorbox}

\subsection{Evaluation Prompt}
\label{sec:eval_prompt}
\begin{tcolorbox}[colframe=black, colback=white, width=\linewidth, boxrule=0.5mm, arc=0mm, sharp corners]
\ttfamily
    
        You are a software quality engineer, your job is to evaluate comments in code according to a rubric.\\

        Please evaluate each comment in the provided \{SOURCE\_LANGUAGE\} code based on the following criteria:\\
        Completeness - Does the comment address all capabilities of the relevant source code? Are significant considerations or functionality omitted?\\
        Hallucination - Does the comment provide true information? Does the description accurately describe code behavior?\\
        Readability - Is the comment clear to read? Is it formatted or phrased in a confusing way?\\
        Usefulness - Is the comment useful? Would it aid a maintainer in understanding the code's functionality, or does it simply "state the obvious" with no added insight?\\
        
        Look through the code and find each individual comment, they will be deliniated by <BLOCK\_COMMENT id> or <INLINE\_COMMENT id> where "id" is an 8-character UUID for the comment that follows.\\
        
        Each comment should be evaluated independently based on the above criteria. Your response should be formatted as a list of JSON objects, with each object corresponding to one comment. Each object should include five keys: `comment\_id`, `completeness`, `hallucination`, `readability`, and `usefulness`. `comment\_id` should have a string value that holds the 8-character UUID associated with the comment. The other four values should each be a JSON object with two keys: `reasoning` (a clear explanation of why the criteria is rated the way it is) and `score` (an integer grade from 0 to 100).\\
        
        Be discerning in your evaluation; only very high-quality comments should be graded at 80 or higher, and 100s should be reserved for exceptionally illuminating documentation. Be a hard grader! If a comment is rated low, be thorough and detailed in your explanation of your score, so that the developer can improve in the future.\\
        
        Below is an example output for a snippet of code with three labeled comments:\\
        \{EXAMPLE\_OUTPUT\}\\
        
        Evaluate the following code:\\
        \{SOURCE\_CODE\}\\

        Don't forget to include your final scores in JSON format!
\end{tcolorbox}

\end{appendix}

\end{document}